\documentclass[amsmath,amssymb,superscriptaddress,notitlepage,twocolumn,nofootinbib]{revtex4-1}
\pdfoutput=1

\usepackage{graphicx}
\usepackage{dcolumn}
\usepackage{bm}
\usepackage{amssymb}
\usepackage{amsmath}
\usepackage{color}
\usepackage[utf8]{inputenc}
\usepackage{comment}
\usepackage{hhline}
\usepackage{float}
\usepackage{multirow}

\def\be{\begin{equation}}
\def\ee{\end{equation}}
\def\ba#1\ea{\begin{align}#1\end{align}}

\newcommand{\reffig}[1]{figure~\ref{fig:#1}}
\newcommand{\refFig}[1]{Figure~\ref{fig:#1}}
\newcommand{\reftab}[1]{Table~\ref{tab:#1}}

\newcommand{\dz}{\delta z_{0.5}^{\rm rec}}
\newcommand{\Dz}{\Delta z^{\rm rec}}

\renewcommand{\v}[1]{\mathbf{#1}}

\newcommand{\rd}{r_{\rm drag}}
\newcommand{\Hunit}{~{\rm km\ s}^{-1}\ {\rm Mpc}^{-1}}
\newcommand{\stlcdm}{sta. recomb.}
\newcommand{\mrlcdm}{mod. recomb.}
\newcommand{\cO}{\mathcal{O}}

\definecolor{ultramarine}{rgb}{0.07, 0.04, 0.56}
\definecolor{cadmiumgreen}{rgb}{0.0, 0.42, 0.24}
\definecolor{indigo(dye)}{rgb}{0.0, 0.25, 0.42}
\definecolor{purple}{rgb}{0.75, 0.0, 1.0}
\usepackage[linktocpage=true]{hyperref}
\hypersetup{
colorlinks=true,
citecolor=ultramarine,
linkcolor=cadmiumgreen,
urlcolor=indigo(dye),
pdfauthor={},
pdftitle={},
pdfsubject={}
}

\begin{document}

\title{Inferences of $H_0$ in presence of a non-standard
  recombination}

\author{Chi-Ting Chiang}
\affiliation{Physics Department, Brookhaven National Laboratory, Upton, NY 11973, USA}

\author{An\v{z}e Slosar}
\affiliation{Physics Department, Brookhaven National Laboratory, Upton, NY 11973, USA}

\date{\today}

\begin{abstract}
  Measurements of the Hubble parameter from the distance ladder are in
  tension with indirect measurements based on the cosmic microwave
  background (CMB) data and the inverse distance ladder measurements
  at 3-4 $\sigma$ level. We consider phenomenological modification to
  the timing and width of the recombination process and show that they
  can significantly affect this tension. This possibility is appealing,
  because such modification affects both the distance to the last
  scattering surface and the calibration of the baryon acoustic
  oscillations (BAO) ruler. Moreover, because only a very small
  fraction of the most energetic photons keep the early universe in
  the plasma state, it is possible that such modification could occur
  without affecting the energy density budget of the universe or being
  incompatible with the very tight limits on the departure from the
  black-body spectrum of CMB. In particular, we find that under this
  simplified model, with a conservative subset of Planck data alone,
  $H_0=73.44_{-6.77}^{+5.50}\Hunit$ and in combination with BAO data
  $H_0=68.86_{-1.35}^{+1.31}\Hunit$, decreasing the tension to $\sim 2\sigma$
  level. However, when combined with Planck lensing reconstruction
  and high-$\ell$ polarization data, the tension climbs back to $\sim 2.7\sigma$,
  despite the uncertainty on non-ladder $H_0$ measurement more than
  doubling.
\end{abstract}

\maketitle

\section{Introduction}
The tension between the local measurement of the Hubble parameter
$H_0$ based on distance ladder and indirectly derived values of Hubble
parameter from higher redshift datasets is a particularly interesting
tension, because both measurements seem to have passed many internal
consistency checks (see
e.g. Refs.~\cite{Cardona:2016ems,Zhang:2017aqn,Feeney:2017sgx,Follin:2017ljs}
for re-analysis of the local distance ladder and
Refs.~\cite{Spergel:2013rxa,Addison:2015wyg,Aghanim:2016sns} for
Planck consistency) and would naturally point to a physics beyond the
standard $\Lambda$CDM model \cite{Bernal:2016gxb}.

The latest distance ladder measurements \cite{Riess:2018byc} observed
the bright Cepheids in both optical and near-infrared to mitigate
saturation as well as to reduce pixel-to-pixel calibration errors, and
the distances of the Cepheids are measured by the Gaia DR2 parallaxes
and HST photometry. This gives $H_0=73.52\pm1.62\Hunit$ (we shall
refer to this as Riess measurement), which is in tension with the
Planck 2018 cosmic microwave background (CMB) data that give
$H_0=67.36\pm0.54\Hunit$ from all Planck measurements combined
\cite{Aghanim:2018eyx}, under the assumption of $\Lambda$CDM
model. More importantly, this tension is robust to a few obvious
resolutions. First, as shown in several papers
\cite{Louis:2014aua,Larson:2014roa,Hou:2017smn,Aylor:2017haa,Huang:2018xle,Aylor:2018drw},
it is robust to change of the CMB data: replacing Planck with a
combination of WMAP and finer resolution experiments gives the same
results. Second, it is robust to very-low redshift modifications to
expansion history (caused, for example, by a rapidly evolving equation
of state of the dark energy): supernovae Ia data can be used to
translate the baryon acoustic oscillations (BAO) determination of
Hubble parameter at $z\sim0.5$ into a $z=0$ measurement under rather
general assumption about smoothness of the evolution of dark energy
component, a method also known as an inverse distance ladder. Results
from this method are competitive and consistent with $\Lambda$CDM
determination using Planck alone. Third, it is also robust to removing
CMB data altogether: calibrating BAO as a standard ruler assuming
standard pre-recombination physics, but using just the big-bang
nucleosynthesis (BBN) measurement of the baryon density. This does not
relieve this tension very significantly either \cite{Addison:2017fdm}.

Therefore, any solution to this problem must involve two ingredients:
i) it must modify the early universe calibration of the BAO ruler and
ii) it must do so while maintaining the excruciatingly precise
measurements of the CMB power spectrum. One possible solution would be
a change in the radiation content of the early universe and/or adding
massive sterile neutrinos.  This improves the fit somehow, but does
not resolve the tension, as the $N_{\rm eff}$ required is inconsistent
with the damping tail measurements of the CMB \cite{Leistedt:2014sia}.

\section{Modification to recombination}
Here we propose an alternative solution: modifying the recombination
history of the universe.  Changing recombination would change the
sound horizon scale of photon-baryon plasma. While the observed angle
of the sound horizon at the photon decoupling, $\theta_s$, remains
the same, different sound horizon requires a different angular diameter
distance to the last scattering surface. Hence, the expansion history
as well as the Hubble parameter will be influenced. In addition, using
BAO as a standard ruler requires the determination of the comoving sound
horizon at the end of the baryonic-drag epoch, $\rd$, from CMB, so
modifying the recombination would also affect the BAO constraint on
the Hubble parameter.

\begin{figure}[h]
\centering
\includegraphics[width=\linewidth]{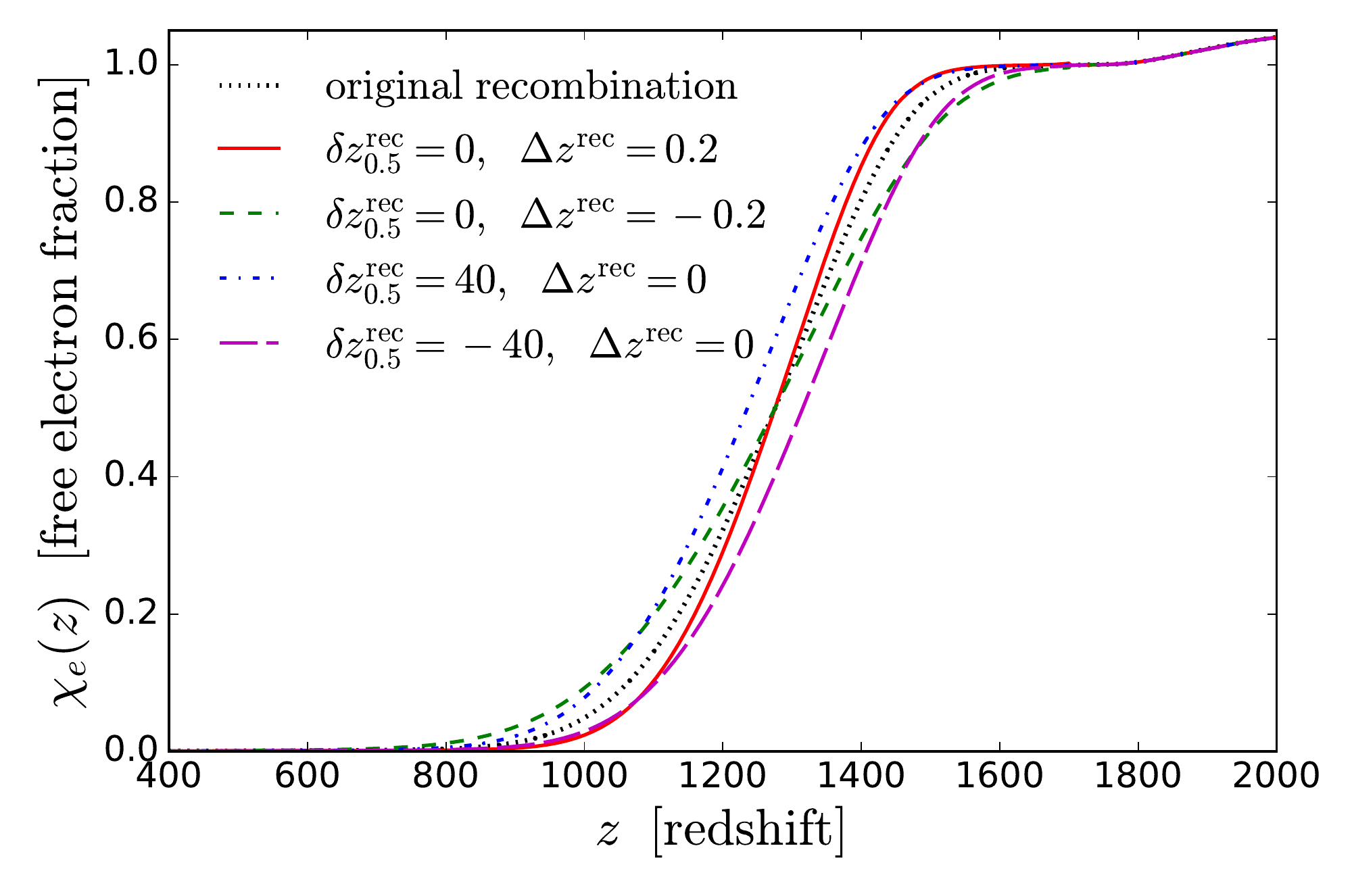}
\caption{The free electron fraction for various values of recombination
parameters $\dz$ and $\Dz$, where $\dz$ shifts the recombination timing
and $\Dz$ changes the recombination width (see the main text for detailed
description of our phenomenological model). In our convention, in terms
of redshift, positive $\dz$ corresponds to a later recombination timing,
and positive $\Dz$ corresponds to a narrower recombination width.}
\label{fig:xe}
\end{figure}

We consider a general phenomenological model that modifies the timing
and width of the recombination, and we apply the changes to
\texttt{CLASS} \cite{Blas:2011rf}. Specifically, when \texttt{CLASS}
computes the free electron fraction $\chi_e$ during recombination, we
first numerically locate the redshift $z_{0.5}^{\rm rec}$ such that
$\chi_e(z_{0.5}^{\rm rec})=0.5$.  The value $z_{0.5}^{\rm rec}$ is
cosmology dependent. We then introduce a mapping redshift
\be
 \tilde{z}=z_{0.5}^{\rm rec}+(z-z_{0.5}^{\rm rec}-\dz)/(1+\Dz) \,,
\ee
where $\dz$ and $\Dz$ parametrize respectively the shift in
recombination timing and the change in recombination width. Finally,
we compute the new free electron fraction
$\tilde{\chi}_e(z)=\chi_e(\tilde{z})$ within $500\le z\le1700$ by
interpolation and set $\tilde{\chi}_e(z)=\chi_e(z)$ outside this
range. By construction, for $\dz=\Dz=0$, we have $\tilde{z}=z$ and the
standard recombination is recovered. In our convention, positive
$\dz$ corresponds to a later recombination timing, and positive
$\Dz$ corresponds to a narrower recombination width. The actual
value of $\rd$ would depend on the other parameters. We vary the
values for the redshift range for modifying the recombination
history, and find that the parameter constraints using CMB and BAO
data are insensitive to the choice, so we stick with the reported
values. \refFig{xe} illustrates the free electron fraction for various
recombination parameters for the Planck 2015 fiducial cosmology \cite{Ade:2015xua}.
A similar exercise has been performed in Ref.~\cite{Hadzhiyska:2018mwh},
but without the shift parameter. Authors find that CMB data alone
gives a tight constraint on the width of recombination process.
Another related work has also been done in Ref.~\cite{Aylor:2018drw}
to measure the sound horizon using the distance ladder calibration.

\section{Results}
To study the impact of the modification to the recombination on the
parameter constraints, we perform Markov chain Monte Carlo (MCMC)
using \texttt{MontePython} \cite{Audren:2012wb,Brinckmann:2018cvx}
with our modified version of \texttt{CLASS}. We apply the Gelman-Rubin
convergence criteria \cite{Gelman:1992zz} and confirm that the chains
achieve $R-1\le0.03$ for all parameters of interest.
We adopt a flat $\Lambda$CDM cosmology with the minimum neutrino
mass, i.e. $m_\nu=0.06~{\rm eV}$ and $N_{\rm eff}=3.046$, as well as
the primordial Helium fraction consistent with BBN.

We use two choices of CMB datasets based on 2015 Planck data
release \cite{Aghanim:2015xee}: \emph{Planck} uses the conservative
choice of likelihood consisting of \texttt{Planck low-l TTTEEE}
and \texttt{Planck high-l TT}; \emph{Full Planck} is more aggressive
by replacing the high-$\ell$ part with \texttt{Planck high-l TTTEEE}
and adding the lensing reconstruction power spectrum \cite{Ade:2015zua}.
In other words, the \emph{Planck} dataset refers to mainly the temperature
power spectrum aided by the large-scale polarization to break certain
degeneracies (most importantly with the optical depth). The \emph{Full Planck}
on the other hand uses the complete information available, including
weak lensing reconstruction which sets the amplitude of matter fluctuations.

In addition to the CMB data, we consider the BAO measurement from the
completed SDSS-III Baryon Oscillation Spectroscopic Survey
\cite{Alam:2016hwk}, which we refer to as \emph{BAO}. Specifically,
the measurement is done with the anisotropic galaxy clustering with
the reconstruction technique, and is the largest galaxy redshift
survey sample to date hence offers the most stringent
constraint. Since using BAO as a standard ruler requires the knowledge
of $\rd$, modifying the recombination affects $\rd$ and will impact
the parameter constraint from BAO.

Finally, we use the BBN prior from on the value of physical baryon
density parameter $\Omega_b h^2$. We choose to measurements from
Ref.~\cite{Cooke:2016rky,Addison:2017fdm}, giving a simple prior
$100\Omega_bh^2=2.260\pm0.034$. In the past, the BBN has been used to
allow the inverse distance-ladder argument to be applied to the BAO
data without CMB prior. However, in the modified recombination model,
this will not work, since one would need to apply priors on not only
$\Omega_b h^2$, but also to $\dz$ and $\Dz$ parameters.  Instead,
since these parameters are degenerate with $\Omega_b h^2$ in the case
of CMB alone, we will apply this prior to CMB data, where it will have
a regularizing function of bringing the model back closer to $\Lambda$CDM.

We have calculated MCMC chains for four possible combinations of the
two Planck datasets with or without the BAO data and for the standard
$\Lambda$CDM model and the modified recombination model. The modified
recombination model has two more parameters than the standard $\Lambda$CDM
model ($\dz$ and $\Dz$).

We apply the BBN prior by importance sampling the chains without
the BAO data. We show the full set of results in \reftab{results}.

\newcolumntype{C}[1]{>{\centering\let\newline\\\arraybackslash\hspace{0pt}}m{#1}}
\newcommand\tspace{\rule{0pt}{2.6ex}} 
\begin{table*}
  \begin{tabular}{ll|C{2.2cm}C{2.2cm}|C{2.2cm}C{2.2cm}|C{2.2cm}C{2.2cm}}
    & & \emph{Planck}  & \emph{Full Planck} &  \emph{Planck}+\emph{BAO} &
        \emph{Full Planck}+\emph{BAO} & \emph{Planck}+\emph{BBN}  & \emph{Full Planck}+\emph{BBN} \\
    \hline
    \multirow{2}{*}{$H_0$:} &
    \stlcdm & $67.65_{-0.94}^{+0.92}$ & $67.86_{-0.61}^{+0.57}$ & $68.06_{-0.59}^{+0.57}$ & $68.04_{-0.46}^{+0.45}$ & $67.91_{-0.86}^{+0.87}$ & $67.99_{-0.57}^{+0.57}$ \\
    & \mrlcdm & $73.39_{-6.68}^{+5.66}$ & $67.17_{-2.17}^{+2.04}$ & $68.86_{-1.35}^{+1.31}$ & $68.17_{-1.18}^{+1.14}$ & $69.94_{-3.13}^{+2.87}$ & $67.87_{-1.93}^{+1.92}$ \\
    \hline
    \multirow{2}{*}{$\frac{(H_0 - H_0^{\rm Riess})}{\sigma_{\rm tot}}$:} &
     \stlcdm & -3.15 & -3.28 & -3.17 & -3.26 & -3.05 & -3.22 \\
    & \mrlcdm & -0.02 & -2.39 & -2.23 & -2.68 & -1.05 & -2.25 \\
    \hline
      \multirow{2}{*}{$\rd$:} &
     \stlcdm & $147.45_{-0.49}^{+0.48}$ & $147.53_{-0.28}^{+0.30}$ & $147.63_{-0.35}^{+0.36}$ & $147.61_{-0.24}^{+0.25}$ & $147.46_{-0.49}^{+0.48}$ & $147.54_{-0.28}^{+0.30}$ \\
    & \mrlcdm & $144.46_{-3.61}^{+3.60}$ & $148.14_{-1.63}^{+1.69}$ & $146.84_{-1.67}^{+1.62}$ & $147.47_{-1.20}^{+1.19}$ & $146.27_{-2.15}^{+2.11}$ & $147.67_{-1.59}^{+1.51}$ \\
    \hline
    \multirow{2}{*}{$\Omega_{\rm m}$:} &
     \stlcdm & $0.3091_{-0.0131}^{+0.0122}$ & $0.3060_{-0.0079}^{+0.0078}$ & $0.3034_{-0.0077}^{+0.0076}$ & $0.3036_{-0.0061}^{+0.0058}$ & $0.3060_{-0.0123}^{+0.0114}$ & $0.3045_{-0.0077}^{+0.0074}$ \\
    & \mrlcdm & $0.2714_{-0.0453}^{+0.0345}$ & $0.3113_{-0.0174}^{+0.0153}$ & $0.2991_{-0.0090}^{+0.0085}$ & $0.3031_{-0.0084}^{+0.0079}$ & $0.2917_{-0.0233}^{+0.0204}$ & $0.3057_{-0.0151}^{+0.0139}$ \\
    \hline
      \multirow{2}{*}{$S_8$:} &
     \stlcdm & $0.845_{-0.025}^{+0.023}$ & $0.832_{-0.013}^{+0.013}$ & $0.835_{-0.017}^{+0.016}$ & $0.828_{-0.011}^{+0.011}$ & $0.840_{-0.023}^{+0.023}$ & $0.830_{-0.013}^{+0.013}$ \\
     & \mrlcdm & $0.822_{-0.035}^{+0.034}$ & $0.835_{-0.016}^{+0.016}$ & $0.843_{-0.019}^{+0.018}$ & $0.829_{-0.011}^{+0.011}$ & $0.837_{-0.025}^{+0.025}$ & $0.831_{-0.015}^{+0.015}$ \\
    \hline
    \tspace  $\dz$: & \mrlcdm & $-37.58_{-44.07}^{+38.14}$ & $6.19_{-16.92}^{+17.23}$ & $-9.59_{-15.01}^{+14.06}$ & $-1.37_{-11.11}^{+10.90}$ & $-16.64_{-23.73}^{+22.10}$ & $1.16_{-16.96}^{+14.47}$ \\
	\hline
    \tspace $\Dz$: & \mrlcdm & $-0.0627_{-0.0637}^{+0.0550}$ & $0.0054_{-0.0182}^{+0.0174}$ & $-0.0230_{-0.0224}^{+0.0209}$ & $-0.0022_{-0.0121}^{+0.0122}$ & $-0.0316_{-0.0312}^{+0.0299}$ & $-0.0004_{-0.0167}^{+0.0154}$
  \end{tabular}
  \caption{\label{tab:results}. 68\% marginalized constraints for a subset of
	relevant cosmological parameters. $H_0^{\rm Riess}=73.52\pm1.62\Hunit$ is the mean
	value of distance ladder measurements from Ref.~\cite{Riess:2018byc} and
	$\sigma_{\rm tot}$ refers to errors from both measurement added in quadrature.
	$S_8$ is defined as $S_8=\sigma_8 (\Omega_{\rm m}/0.3)^{0.5}$ \cite{Abbott:2017wau}.}
\end{table*}

\begin{figure}[h]
\centering
\includegraphics[width=\linewidth]{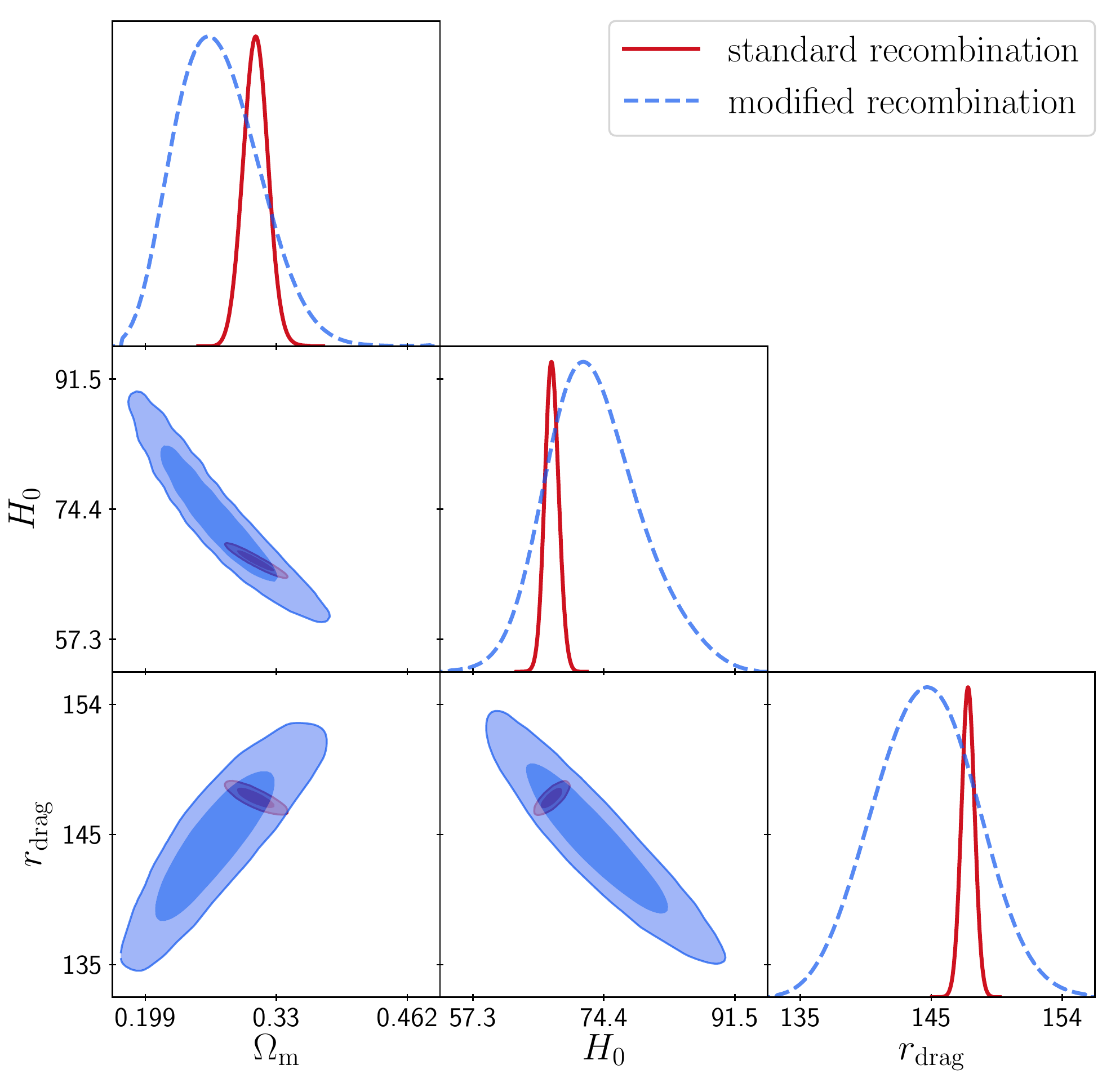}
\caption{Triangle plot of $\Omega_{\rm m}-H_0-\rd$ from \emph{Planck}
dataset alone. The red and blue contours display the parameter
constraints for standard and modified recombination, respectively.
The modified recombination significantly relaxes the $H_0$ constraint
and the $\rd$ calibration of the BAO ruler.}
\label{fig:relax}
\end{figure}

\begin{figure}[h]
\centering
\includegraphics[width=\linewidth]{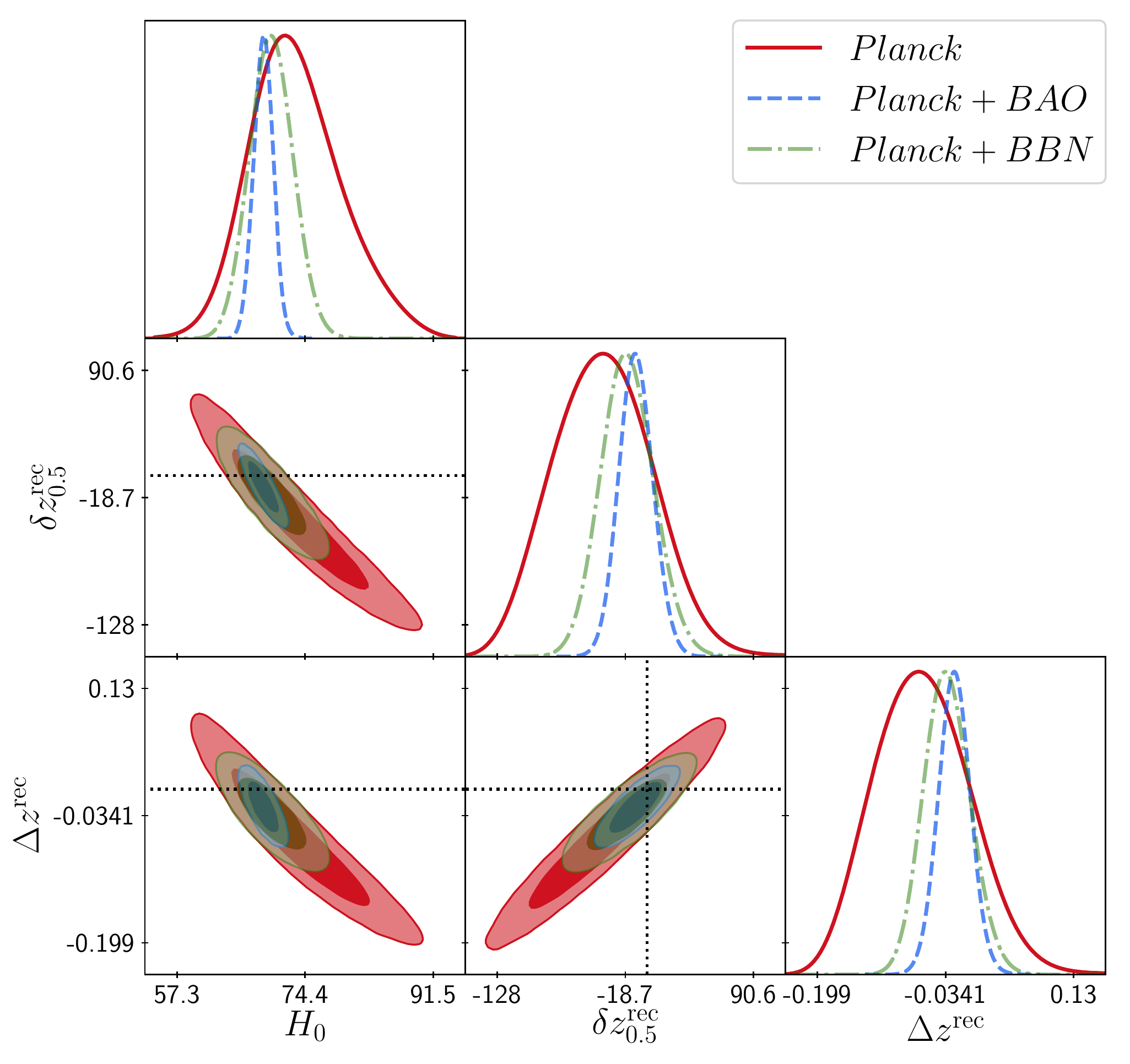}
\caption{Triangle plot of $H_0-\dz-\Dz$ of modified recombination.
The red, blue, and green contours display respectively the parameter
constraints for \emph{Planck}, \emph{Planck}+\emph{BAO}, and
\emph{Planck}+\emph{BBN}. The dotted lines indicate $\dz=0$ or $\Dz=0$.
Adding either BAO or BBN helps break the degeneracies, but the contours
remain significantly enlarged compared to the standard recombination.}
\label{fig:fixrelax}
\end{figure}

\begin{figure}[h]
\centering
\includegraphics[width=\linewidth]{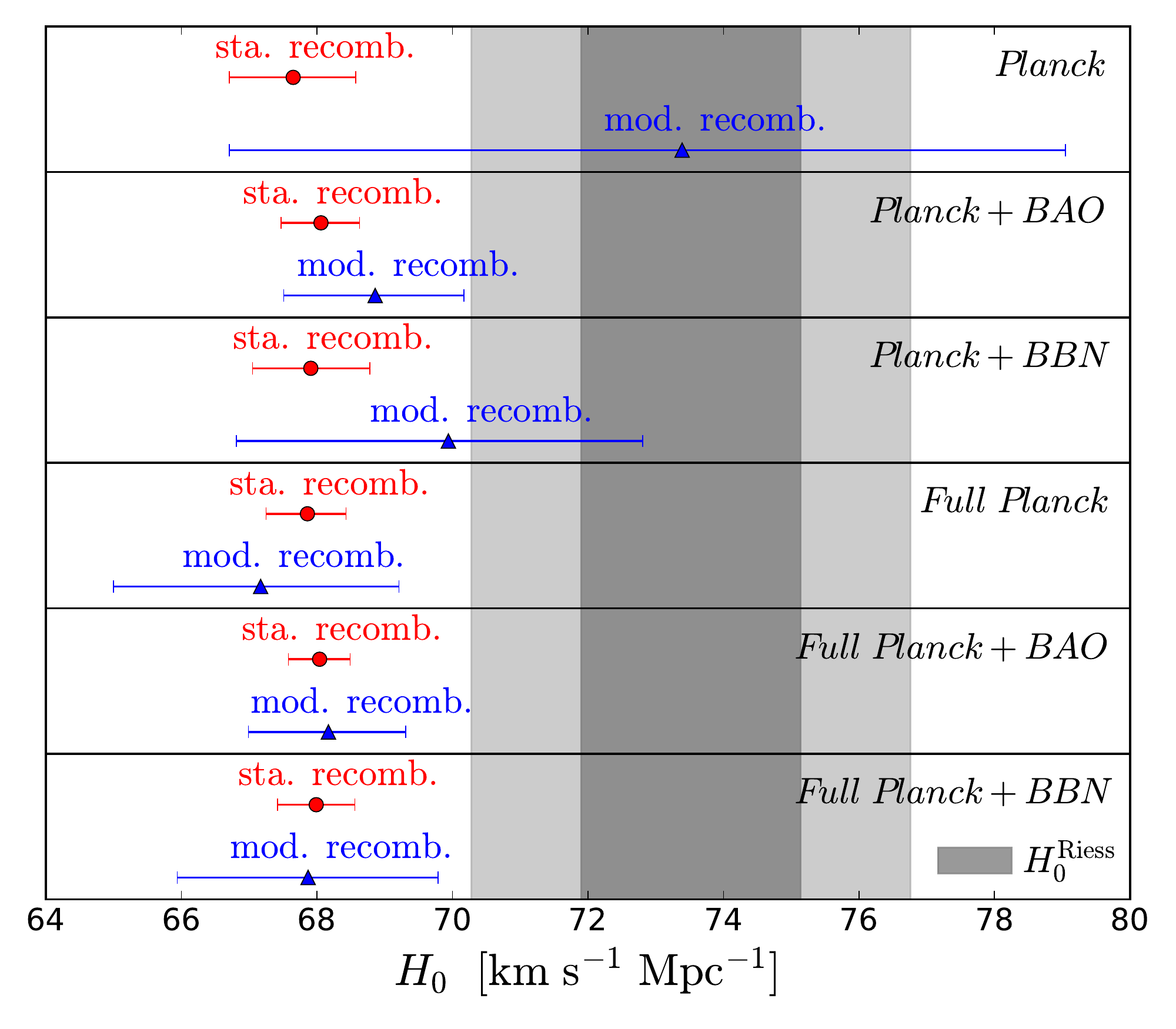}
\caption{Constraints on the Hubble parameter for various data
combinations shown in each panel. The red circles and blue triangles
with error bars display the constraints for the standard and modified
recombination, respectively. The gray band display the 1 and $2\sigma$
constraint from Ref.~\cite{Riess:2018byc}. It is apparent that including
the additional recombination parameters $\dz$ and $\Dz$ significantly
enlarges the constraint on $H_0$, making the constraint from the early
universe more consistent with the local distance ladder measurement.}
\label{fig:money}
\end{figure}

This table contains the main results and we will start by highlighting
some of the main conclusions from this data. \refFig{relax} displays
the triangle plot for \emph{Planck} data alone. We see that the basic
picture holds: by relaxing the physical connection between recombination
timing the constraints on the basic $\Lambda$CDM parameters relax.
The relaxation is not complete, because a recombination occurring at a
different redshift would result in a radically different power spectrum.
Nevertheless, the parameter constraints inflate prominently. In particular,
we see that the timing of recombination significantly affects the implied
constraint on $\rd$, the ruler size for BAO. Most importantly, the error
on $H_0$ increases by a factor of 7 with a central value that is delightfully
close to the Riess measurement. The main question is whether this survives
the addition of other datasets.

In \reffig{fixrelax} we see that adding other data to \emph{Planck}
dataset helps bring some of the parameters back closer to their
fiducial values. We find that \emph{BBN} constraint helps in this
case, but the errors on $H_0$ remain large. The \emph{BAO} data,
however, pull $H_0$ back to the fiducial value, albeit with a
$1\sigma$ shift in the central value and a factor of $\sim 2$
increase in error bars. The tension with the Riess measurement
is relaxed by about one standard deviation and remains above $2\sigma$.

The results on $H_0$ constraints are summarized in \reffig{money}.
The additional recombination parameters $\dz$ and $\Dz$ significantly
enlarges the constraint compared to the standard recombination, making
the $H_0$ measurement from the early universe more consistent with
the distance ladder measurement. In particular, for the \emph{Planck}
data combinations the $H_0$ central values are shifted to a higher
values, closer to the Riess measurement. However, adding Planck
lensing reconstruction and high-$\ell$ polarization data pulls
$H_0$ back to the values of standard recombination. Nevertheless,
the increase of uncertainty still relieves the tension with the
distance ladder measurement.

\section{Discussion}

In this paper we have shown that there exist a phenomenological
expansion of the $\Lambda$CDM model that is capable of significantly
decreasing the tension between distance ladder Hubble parameter
measurements and indirect determinations relying on early universe
physics. The main feature of this model is that it simultaneously
affects the standard CMB parameter fitting, while also changing the
BAO ruler calibration.

When considering the \emph{Planck} data alone, the new parameters bring
degeneracies that relieve the tension completely. However, the additional
data, such as \emph{BAO} data and the \emph{Full Planck} data result in
tension reappearing, but is now at $2-3\sigma$ level rather than $3-4\sigma$
level.

For completeness, we also calculate the effect on other parameters,
such as the matter density and the lensing amplitude parameter $S_8$
and find that they are not significantly affected.

Importance sampling our most constrained parameter combination of
\emph{Full Planck}+\emph{BAO} with the $H_0$ prior from the Riess
measurement, we find the marginalized values of
$\dz=-17.07_{-9.91}^{+8.97}$ and $\Dz=-0.0183_{-0.0111}^{+0.0100}$.
This indicates that a higher $H_0$ values favors an earlier and wider
recombination in terms of redshift, and the exact value of $\rd$ will
depend on the other parameters. The same qualitative trend is seen
for using \emph{Planck} dataset alone, but the constraints on the
modified recombination parameters are statistically consistent with
zero.

In this exploratory paper we do not propose a physical model that
could implement this idea. Nevertheless, we make some relevant
remarks. An important point is that due to the large over-abundance
of photons compared to baryons, with the photo-to-baryon ratio of
$n_\gamma/n_b \sim 6\times 10^{-10}$, only a very small fraction of
photons deep in the exponential suppression of the distribution
participate in keeping the universe ionized up to recombination. These
photos are today redshifted to frequencies $\nu \gtrsim 1.5~$THz and
are therefore not constrained by the precise FIRAS measurement of CMB
black-body spectrum \cite{Mather:1993ij,Fixsen:1996nj}. Second, while
these photons an order of magnitude more energetic compared to
typical CMB photons, they contribute negligibly to the energy density
of radiation fluid simply because they are so few of them (by
construction, their total number is suppressed by the baryon
fraction). Therefore, it is entirely possible to have $\cO(1)$
modifications to the distribution of these photons, without disturbing
the energetics of the universe at all. If the modification is the
high-energy tail of the CMB spectrum, we have shown that this
solutions requires \emph{fewer} photons in the tail in order to
recombine at a higher redshift.  Future CMB spectral distortion
observations can help determine the recombination physics more
accurately, further testing the validity of resolving the $H_0$
tension by a non-standard recombination.

Finally, we conclude by saying that we did not obsess over the details
of model comparison, because it is premature. We reiterate that this
model is completely phenomenological and that a more concrete, physically
based model could provide better fits for the union of data. Presumably,
any model grounded in actual physics will have a somewhat more complicated
properties that are not well described by a simple shift and stretch
parameterization. Nonetheless, we have demonstrated that should the
data continue to show an ever increasing tension in the two determinations
of $H_0$, that modified recombination offers a possible reconciliation mechanism.

\section*{Acknowledgments }

AS acknowledges fruitful discussions with Antony Lewis which have
eventually led to the idea presented in this paper.

\bibliography{ms}

\end{document}